\documentclass[prl,twocolumn,letterpaper,showpacs]{revtex4-1}
\usepackage{graphicx}
\usepackage{color}
\usepackage{amssymb}
\usepackage{amsmath}
\usepackage{amsfonts}

\begin{document}

\title{Experimental evidence of solitary wave interaction in Hertzian chains}

\author{Francisco Santibanez}
\email[Electronic address: ]{francisco.santibanez@usach.cl}
\author{Romina Munoz}
\author{Aude Caussarieu$^{\dagger}$}
\author{St\'ephane Job$^{\dagger\dagger}$}
\author{Francisco Melo.}
\affiliation{Departamento de F\'{\i}sica, Universidad de Santiago de Chile, Avenida Ecuador 3493, Casilla 307, Correo 2, Santiago de Chile.}
\affiliation{$^{\dagger}$Laboratoire de Physique, ENS de Lyon, 46, all\'ee d'Italie F69007 Lyon, France.}
\affiliation{$^{\dagger\dagger}$Supmeca, 3 rue Fernand Hainaut, 93407 Saint-Ouen Cedex, France.}

\date{\today}


\begin{abstract}
We study experimentally the interaction between two  solitary waves that approach one to another in a linear chain of spheres interacting via the Hertz potential. When these counter propagating waves collide, they cross each other and a phase shift respect to the noninteracting waves is introduced, as a result of the nonlinear interaction potential. This observation is well reproduced by our numerical simulations and it is shown to be independent of viscoelastic
 dissipation at the beads contact.  In addition, when the collision  of equal amplitude and synchronized counter propagating
 waves takes place, we observe that two secondary solitary waves emerge from the interacting region. The amplitude of secondary solitary waves is proportional
 to the amplitude of incident waves.  However, secondary solitary waves are stronger when the collision occurs at the middle contact
  in chains with even number of beads.  Although numerical simulations correctly predict the existence of these waves, experiments show that
   their respective amplitude are significantly larger than predicted.  We attribute this discrepancy to the rolling friction at the beads
   contacts during solitary wave propagation.
\end{abstract}


\maketitle
\section{\label{sec:intro} Introduction}
As it was shown in the pioneering work of V. Nesterenko~\cite{book:dynamics_nesterenko,art:nesterenko_propagation,art:nesterenko_lazaridi}, wave propagation in one dimensional assembly of beads has a rich content in physics due to the nonlinear nature of the interaction between spheres. If the alignment does not support any static load, linear acoustic modes of propagation (phonons) are forbidden and any infinitesimal small perturbation propagates as solitary waves. This fully nonlinear regime corresponds to a {\it sonic vacuum}~\cite{book:dynamics_nesterenko}. Several studies have reported both experimentally and numerically a large number of phenomena associated with this type of waves in Hertzian linear chains, ranging from solitary wave interaction with boundaries~\cite{art:fauve_coste_1997,art:Job_melo_2005,art:nesterenko_daraio_2005,art:job_melo_2007} and wave mitigation in tapered chains~\cite{art:job_melo_2006,art:doney_sen_2005,art:harbola_pulse_propagation_2009,art:harbola_pulse_propagation_2010}, to energy localization in chains with mass defects ~\cite{art:job_santibanez_2009,art:Boechler_discrete_2010,art:Theocharis_Intrinsic_2010} and modification of wave propagation in lubricated granular chains~\cite{art:herbold_nesterenko_2005,art:job_santibanez_2008}. The weakly nonlinear regime, for which the amplitude of a stress perturbation is smaller but comparable to the imposed static stress, also exhibits well known characteristics. For instance, it has been shown that the dynamics of chain of weakly loaded spheres obey the Korteweg- de Vries equation ~\cite{book:dynamics_nesterenko,book:drazin_johnson_2002,book:peyrard_dauxois_2004}, which is known in turn to support solitons, kinks and breathers as exact solutions ~\cite{art:manciu_sen_2000}. One of the remarkable features of the weakly nonlinear KdV solitons resides in their ability to interact nonlinearly with each other and escape a collision by conserving not only their momentum and energy, but also the full characteristics of the initial waves (shape, width, amplitude…). The only trace of the nonlinear nature of the interaction is a slight post-collision phase shift of each wave, compared to the non-interacting configuration~\cite{book:drazin_johnson_2002,book:peyrard_dauxois_2004}. One of the main features of solitary wave is the nonlinear character of their interaction for which the linear superposition principle does not hold. Here, we investigate the interaction of two identical solitary waves traveling in opposite directions in a linear chain of spherical elastic beads. In the nonlinear work frame, the amplitude of the total wave during the interaction is larger than the simple sum of two individual wave.  Since the wave speed is a nonlinear function of wave amplitude, when the interaction ends, both waves have covered the distance where they interacted in a shorter time ~\cite{art:manciu_sen_2000}. The later implies the existence of a phase-shift in the form of a time delay, as for KdV solitons in the weakly nonlinear regime. The observation of such a delay thus may serve to probe the exact nature of the strongly nonlinear solitary waves that propagate in one-dimensional weakly loaded granular structures. It has also been discussed  that energy partition in chains - for counter propagating waves initiated at the borders at equal time and equal amplitude - with a even number of beads (where waves collide at the beads contact), differs from the case of a chain with odd number of beads (where the bead at the center does not move during the wave interaction). In the first case, the interaction generates an excess of potential energy at the bead's contacts which later contributes to the formation of secondary solitary waves (SSW)~\cite{art:avalos_doney_2007,art:avalos_sen_2009}. The study of frontal interaction between solitary waves of significantly different amplitudes and of over-taking solitary waves traveling in the same direction~\cite{art:zhen_shun_2007} both require setups having an important number of beads. In the present work, only frontal collision between waves of similar amplitude were considered, which has the advantage of dealing with short chains and accordingly small dissipation. This paper is organized as follows: a brief description of the soliton-like solution for a one dimensional granular chain is first reminded, a full description of our experimental setup and observations is then given, and an analysis of experimental and numerical results are finally discussed in terms of solitary wave interaction time.

\section{\label{sec:analysis} Analysis and Numerical Simulations}	
The physical behavior of a chain of \emph{N} beads under elastic deformation is described by using the pairs Hertz potential~\cite{book:landau_elasticidad}, $U_H = (2/5)\kappa \delta^{5/2}$ where $\delta$ is the deformation of two consecutive spheres, $\kappa^{-1} = (\theta + \theta')(R^{-1} + R'^{-1})^{1/2}$, $\theta = 3(1-\nu^2)/4Y$ and $R$ and $R'$ are the radii of curvature at the contact. $Y$ and $\nu$ are the Young modulus and Poisson's ratio, respectively. Since the force felt at the contact is  the derivative of the potential energy with respect to $\delta$ ($F_H = \kappa \delta^{3/2}$ , and $F_H = 0$ if the beads lose contact), it is possible to describe the dynamics of the chain by the following nonlinear coupled differential equations:\\
\begin{equation}
m \ddot{u}_n = \kappa \left[ (u_{n-1} - u_n)^{3/2} - (u_n - u_{n+1})^{3/2} \right].
\label{eq:eq01}
\end{equation}
Here, time derivatives are denoted by overdots and \emph{m} is the mass of each bead. Under the long-wavelength approximation $\lambda \gg R$ (where $\lambda$ is the characteristic length of the perturbation) for the case of vanishing external static load, one can obtain the continuum limit of Eq.~(\ref{eq:eq01}) by replacing the discrete function $u_{n \pm 1}(t)$ by the corresponding Taylor expansion of the continuous function $u(x\pm R,t)$. Keeping terms up to the fourth order in spatial derivatives, it is possible to obtain an equation for the strain $\psi = -\partial_x u$
\begin{equation}
\ddot{\psi} \simeq c^2 \partial_{xx} \left[ \psi^{3/2} + (2/5)R^2 \psi^{1/4}\partial_{xx}(\psi^{5/4}) \right]
\label{eq:eq02}
\end{equation}
which admits an exact periodic solution in the form of a traveling wave, $\psi(\xi = x - vt)$, with speed \emph{v}. This solution is:\\
\begin{equation}
\psi = (5/4)^2 (v/c)^4 \cos^4{[\xi/(R\sqrt{10})]}
\label{eq:eq03}
\end{equation}
where $c=(2R)^{5/4}(\kappa/m)^{1/2}$. Although this is a truncated solution of Eq.~(\ref{eq:eq01}), it is well established that one hump $\left[-\pi/2 < \xi/(R\sqrt{10}) < \pi/2 \right]$ of this periodic function represents a solitary wave solution~\cite{book:dynamics_nesterenko}. Finally, approximating the spatial derivative by finite differences, the strain in the chain reads $\psi = \delta/(2R)$. Thus, the force at the interface, $F \simeq \kappa (2R\psi)^{3/2}$,  and the velocity \emph{v} become\\
\begin{equation}
F \simeq F_{m} \cos^6{\left[ \frac{x - vt}{R\sqrt{10}} \right]},
\label{eq:eq04}
\end{equation}
\begin{equation}
v \simeq \left( \frac{6}{5\pi\rho} \right)^{1/2} \left(\frac{F_{m}}{\theta^{2} R^{2}}\right)^{1/6} \simeq \left( \frac{6}{5\pi\rho} \right)^{2/5} \left(\frac{V_{m}}{\theta^{2}}\right)^{1/5}.
\label{eq:eq05}
\end{equation}
The second term expression for the wave velocity is obtained by relating the maximum force $F_{m}$ to the maximum bead velocity $V_{m}$. Experimentally, the maximum force $F_{m}$, the time of flight \emph{T} and the duration $\tau$ are obtained here from fitting the loading part $\left( -\tau < t-T < \tau\right)$ of the measured force as a function of time  to Nesterenko's solution $F(t) = F_{m}\cos^{6}{\left[(t-T)/\tau\right]}$~\cite{art:Job_melo_2005,art:job_santibanez_2008} (with $2\tau = 2R\sqrt{10}/v$).  As a way to complete the experimental information shown in the following sections, we perform numerical simulations by solving Eq.~(\ref{eq:eq01}) for a chain of $N=25$ (or $N=26$) beads, using a fourth order Runge-Kutta algorithm, with a time step smaller than the minimum physical time in the system  and which fulfills the energy conservation up to relative variations of order $10^{-9}$. In our numerical simulations we have taken viscoelastic dissipation into account by using a nonlinear viscoelastic solid model~\cite{art:kuwabara_kono_1987} ($F_{v} = (3/2)\eta \kappa \delta^{1/2}\dot{\delta}$, with viscous relaxation time $\eta=0.58 \mu$s, derived for this set of beads, from the measurement of restitution coefficient~\cite{art:Job_melo_2005}). To initiate the waves, initial velocities are applied over the first and last beads ($+V_{0}$ and $-V_{0}$, correspondingly).  The result of our simulations allow us to estimate the force felt by all contacts as function of time and provide a direct comparison with the forces measured in our experiments.\\
\section{\label{sec:setup} Experimental Setup}
\begin{figure}[h]
\includegraphics[width=.45\textwidth]{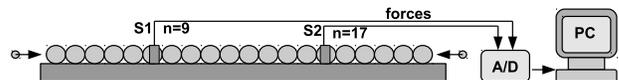}
\caption{\label{fig:fig01} Experimental setup showing a chain of beads with sensors S1, S2 and acquisition facilities. The wave that is initiated at the left-hand side of this diagram and passes first through sensor S1 corresponds to LSTW, the opposite wave is named RSTW.}
\end{figure}
Our experimental setup consists of a linear chain of steel beads.  Beads are {\em 100C6} steel roll bearings with density $\rho=7780$~kg/m$^3$, Young's modulus $Y=203\pm4$~GPa and Poisson ratio $\nu=0.3$. The chain is made of either $N = 26$ or $N =25$ equal beads (radius $R=13$~mm) with free boundaries, as shown in
 Fig.~\ref{fig:fig01}. The beads, barely touching one another, are aligned on a horizontal Plexiglas track which eliminates transversal degrees of freedom. Two solitary waves are initiated from the impact of two small strikers ($R_s=4$~mm) located at each ends of the chain. Strikers are released from a determined height by using
  an electronic circuit that allows control over the time of fall of each striker and consequently, over the synchronization of the entrance of each
  pulse in the chain. The pulses are monitored by measuring the dynamic force with two piezoelectric transducers ({\em PCB 200B02},
  sensitivity $11.24$~mV/N and stiffness $1.9$~kN/$\mu$m), each being inserted inside a bead cut in two parts, as presented in a
  previous work~\cite{art:Job_melo_2005}. The total mass of an active bead and its overall rigidity (led by the contact stiffness with neighbors) both matches the characteristics of an original bead. The
   embedded sensors thus allows non-intrusive measurements of the force inside the chain.

   In order to register solitary
    waves interactions and the further resulting mechanisms, such as  secondary solitary waves generation, both sensors are placed in
    symmetrical positions with respect to the middle of the chain (as shown in Fig.~\ref{fig:fig01}). These positions are
     chosen within two compromises namely, sensors must be far enough from the free boundaries, in order to avoid registering the
     transient formation of the incident solitary waves,  and they must be separated by a distance larger than the typical width of a pulse (about five beads), in order to distinguish the secondary solitary waves well separated from the main pulses.\\
\section{\label{sec:experimental} Experimental Observations}
\begin{figure}[h]
\includegraphics[width=.4\textwidth]{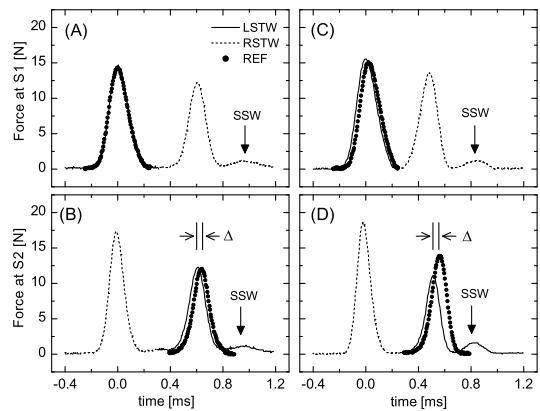}
\caption{\label{fig:fig02} Experimental result showing the time delay for odd and even chains, left column corresponds to a chain of $25$ beads and right column to $26$ beads. Plots (A) and (B) show the force felt by sensors in positions $n=9$ and $n=17$ for odd ($N=25$) chain, while plots (C) and (D) correspond to force in $n=10$ and $n=17$ for the even chain ($N=26$). In all cases, the solitary wave traveling from left to right (LSTW) in our setup is plotted in solid line, dashed line indicates that the wave travels from right to left (RSTW), a reference wave traveling from left to right is shown with dots. (Notice that maximum forces of SW are different for odd and even chains)}
\end{figure}

\begin{figure}[h]
\includegraphics[width=.4\textwidth]{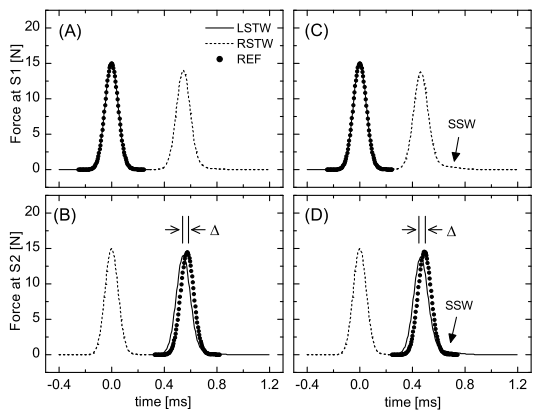}
\caption{\label{fig:fig03} Simulation results  (sensors not included). Same notation used for Fig.~\ref{fig:fig02}. It is observed that viscoelastic dissipation does not completely accounts for the decay in maximum force. Also, SSW appear in plots (C) and (D) with amplitudes much smaller than observed in experiments for the same incident waves.}
\end{figure}

Experimentally, care is taken to establish the exact zone where pulses interact. Sources of errors are mainly due to amplitude differences of the left-hand side traveling wave (LSTW) and right-hand side traveling wave (RSTW) and the presence of small distance gaps between beads which may introduce path differences between waves and lead to an undesirable shift in the interaction zone. The later is controlled by softly and carefully compressing the chain by hand in order to eliminate any distance gap between beads before any acquisition. For each releasing of the striker beads, which are dropped from the same height, a large set of data that consists in the force measurements of incident and passed-through resulting waves, is recorded from both sensors. When the incident waves registered by the sensors had amplitude difference no greater than $5\%$, it is considered that solitary waves have crossed at the middle of the chain (or, at most, at a distance corresponding to one bead's radii from the middle). Data for which this condition is not fulfilled are not retained for further analysis. When this condition is fulfilled, the reference solitary wave is obtained by initiating a single solitary wave with same incident amplitude, by dropping a single striker from the same height. The reference wave travels along the chain from left to right without any interaction. The reference data set contains the passing of the reference solitary wave through both sensors, from which we obtain the unperturbed wave's time of flight, $T_{s}$.\\
In Fig.~\ref{fig:fig02}, both waves are identified as they travel along the chain in opposite directions. Indeed, after the collision, both waves (RSTW and LSTW) are registered in opposite sensors and eventually leave the chain by producing a small gap at the end beads. A time shift appears between the resulting waves and the reference wave, and we observe that the later arrives to the sensor with a time delay $\Delta$, although the reference wave has a larger amplitude (thus greater speed) than the resulting waves after the collision. The measurement of $\Delta$ is independent of the position of the sensors given that it is obtained from the subtraction of the flight times of the perturbed and unperturbed waves, both being measured at the same location. Still, we observed the formation of secondary solitary waves (shown in plots (C) and (D) of Fig.~\ref{fig:fig02}) in chains with either odd or even number of beads. These secondary solitary waves are captured with smaller amplitude in the numerical simulation as shown in Fig.~\ref{fig:fig03}. 
\section{\label{sec:discussion} Discussion}
\begin{figure}[h]
\includegraphics[width=.4\textwidth]{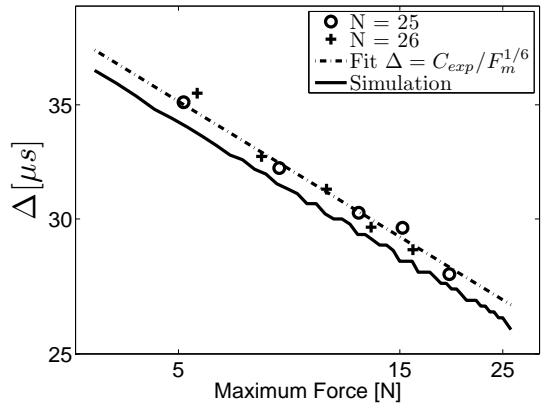}
\caption{\label{fig:fig04} Delay between reference and crossing solitary waves, $\Delta$, as a function of maximum incident wave amplitude, $F_{m}$. Empty bullets and crosses represent experimental data for $N_{o}=25$ and $N_{e}= 26$, chains. Solid line represents the simulation result and dashed line shows the fit of experimental data using $\Delta = C_{exp}F_{m}^{-1/6}$, with $C_{exp} =45.9\pm0.5$[$\mu$sN$^{1/6}$] being in satisfactory agreement with the ballistic approximation prediction $C_{bal} = 49.74$[$\mu$s$N^{1/6}$] (see Eq.~\ref{eq:eq06}).}
\end{figure}

As shown in Fig.~\ref{fig:fig04}, the time delay between a collided wave and the reference is proportional to $F_m^{-1/6}$. This observation indicates that the effective interaction between solitary waves is a function proportional to the duration of solitary waves, $\tau$. By considering a reference frame over one of the incident solitary waves, the interaction itself can be seen as if this wave passes over a localized compression zone inside the chain. This local change in static load produces an increase in the amplitude of the wave and thus an increase of its velocity.  However, the numeric realization of this idea requires the introduction of an estimation of a characteristic interaction distance, which might lie in the range of the typical width of a pulse, $2R\sqrt{10}$.  Another more consistent alternative view is to consider the incident waves as two beads approaching with velocity $2V_m$ one to another~\cite{art:job_melo_2007}. As before, $V_m$ is the maximum velocity of the beads during the solitary wave propagation.  Since the collision time is a function of relative velocity between beads, the time delay is simply given by the collision time of a single bead impinging a static one with speed $V_m$, minus the collision time for a frontal collision where beads have relative velocity $2V_m$ ~\cite{book:landau_elasticidad}.  This is,\\

\begin{eqnarray}
\Delta  &=&  \frac{2.94}{V_{m}^{1/5}} \left(\frac{5 \pi \rho R^3 }{3\kappa}\right)^{2/5} \left[ 1-\frac{1}{2^{1/5}} \right] \nonumber \\
        &=&  \left[ 0.7612 \left(\frac{5 \pi  \rho}{6}\right)^{1/2} \left( \theta R^{4} \right)^{1/3} \right] F_{m}^{-1/6} \nonumber \\
        &=&  C_{bal} \times F_{m}^{-1/6}
\label{eq:eq06}
\end{eqnarray}
Here the  pre-factor, $C_{bal}=49.74$[$\mu$sN$^{1/6}$], is obtained explicitly from Eq.~(\ref{eq:eq05}) for the ballistic collision approximation. This results is constructed under the assumption of elastic interaction including the viscoelastic dissipation effect. It is worth mentioning that viscoelastic dissipation does not contribute to the time shift because such dissipation increases the interaction duration of two colliding beads in a factor $\eta$ independently of the intensity of the impacting velocity. Thus, this effect cancels out when estimating the time shift. This result is well verified in our numerical calculations. However, in Fig.~\ref{fig:fig04} we observe a small difference between experiments and simulations for the measured time shift. This difference corresponds to the increase in wave duration due to friction dissipation between the plexiglas track and beads. This effect implies that for a given incident force, pulse duration is slightly longer in experiments than in simulations, the actual interaction between waves in experiments is thus longer than in simulations.

\begin{figure}[h]
\includegraphics[width=.4\textwidth]{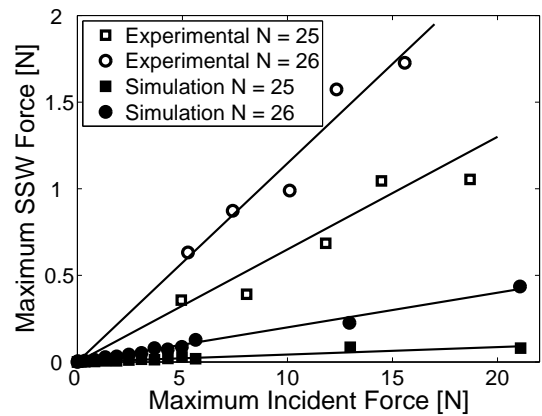}
\caption{\label{fig:fig05} Amplitude of the SSW as function of incident wave amplitude. Empty markers correspond to experimental data while solid markers show simulation results.}
\end{figure}

Fig.~\ref{fig:fig05} shows the SSW amplitudes as function of incident wave amplitude.  The amplitude of SSW is proportional to the amplitude of incident waves.  However, secondary solitary waves are stronger when the collision occurs at the middle contact in chains with even number of beads. This behavior has been predicted by Sen and coworkers \cite{art:manciu_sen_2000} who showed that SSW are generated by a subsequent release of a little unbalance of potential energy at the region where interaction takes place. For the case of solitary waves interacting at the contact between two beads, the energy unbalance is larger.

 Although our numerical simulations correctly predict the existence of these waves, in qualitative agreement with results in Ref. \cite{art:manciu_sen_2000},  experiments show that SSW amplitudes are significantly larger than predicted. This difference might have several sources, one of these is the energy dissipation that introduces a new time scale in the system.  However, a first mechanism of dissipation is due to viscoelastic behavior of the bead material which is included in simulation and does not explain the large amplitude of SSW.  The second dissipation mechanism is due to friction between the plexiglas track and the bead, the contribution of friction also affects the duration and amplitude of the wave. However, as it was shown in previous work~\cite{art:Job_melo_2005}, this effect is of the same order than viscous dissipation in our setup.  Thus, neither viscous nor frictional dissipation can account for the large amplitude SSW observed in experiments.\\

It is worth mentioning that recent numerical achievements revealed that the presence of a viscous contribution, proportional to the relative beads velocity~\cite{art:rosas_romero_2007,art:rosas_romero_2008} (as it would be for instance in presence of a purely viscous fluid between beads~\cite{art:job_santibanez_2008,art:herbold_nesterenko_2005}) or in the form of heuristic term proportional to some power of the absolute value of the relative beads velocity~\cite{art:carretero_khatri_2009}, in addition to the nonlinear Hertz potential, may induces spontaneous secondary waves as a result of a slow relaxation of the viscous stress. However, the typical time scale of the  pulses~\cite{art:rosas_romero_2007,art:rosas_romero_2008} noticeably differs from the leading pulse, which were not observed in our experiments. The model presented in~\cite{art:carretero_khatri_2009} does not conserve momentum by construction and might reveal in turn the existence of a complex mechanism involving losses of momentum through a frictional coupling between the beads and external walls or supports.

 In fact, we suspect the difference between the observed SSW amplitude and the numerical prediction to be due to small rotation of the beads~\cite{art:merkel_Tournat_2010} induced by the friction between the beads and the track.  The friction and the rotational degrees of freedom are not included in our model, despite one can expect that when a single solitary wave propagates in the chain of beads, the friction on the track induces a small thwarted rotation of each bead. In the case where two solitary waves traveling in opposite directions interact in one contact, both adjacent beads partially roll in over each other, which ultimately leads to an excess of potential energy stored into compressional deformation.  In the other case, when the interaction occurs at the center of one bead, both adjacent beads roll over a static one, which leads to shear deformation of the contacts of the static bead. In both cases, the elastic energy stored through these additional mechanisms might be released afterward and induces SSWs stronger than from the frictionless prediction.

\section{\label{sec:conclusions} Conclusions}
In conclusion, we have been able to measure the phase shift produced by the frontal collision of two equal
solitary waves in one dimensional Hertzian chains. We compared our experimental observations  to the
 numerical solution of Eq.~(\ref{eq:eq01}).  Moreover, by using a well known result from linear elasticity theory, we were able to obtain a
 quantitative agreement to account for the time duration of the interaction between two colliding spheres with relative
 velocity $2V_{m}$ and contrast it to the unperturbed propagation where spheres collide with relative velocity $V_{m}$.
 Secondary solitary waves have been detected for chains with odd and even number of beads. The amplitude difference between experimental
  SSWs and calculated ones has been interpreted as  consequence of an additional potential energy storage due to of solid friction at the contact beads.
\begin{acknowledgments}
This work was supported by Conicyt Chile and ANR France under research program ANR-Conicyt $N^{o}$ 011.
\end{acknowledgments}


\end{document}